\documentclass[aps,amsmath,amssymb,prl,twocolumn,superscriptaddress,showpacs]{revtex4}

\usepackage{bm}
\usepackage{epsfig}
\usepackage[usenames]{color}
\usepackage{graphicx}

\newcommand{\el}{{\rm el}}
\newcommand{\inel}{{\rm in}}
\newcommand{\SFM}{{\rm SFM}}
\newcommand{\SPM}{{\rm SPM}}
\newcommand{\opt}{{\rm opt}}
\renewcommand{\P}{{\cal P}}
\newcommand{\T}{{\cal T}}
\newcommand{\E}{{\cal E}}

\newcommand{\av}[1]{\left\langle #1\right\rangle}

\renewcommand{\paragraph}[1]{\textit{#1.---} } 
\newcommand{\paper}{paper }

\newcommand{\etal}{{\it et al.}}

\begin{document}

\title{Electron Transport in Nanogranular Ferromagnets}
\author{I.~S.~Beloborodov}
\affiliation{Materials Science Division, Argonne National Laboratory, Argonne, Illinois 60439, USA}
\affiliation{James Franck Institute, University of Chicago, Chicago, Illinois 60637, USA}
\author{A.~Glatz}\author{V.~M.~Vinokur}
\affiliation{Materials Science Division, Argonne National Laboratory, Argonne, Illinois 60439, USA}

\date{\today}
\pacs{71.10.-w, 75.10.-b, 73.43.Qt}

\begin{abstract}
We study electronic transport properties of ferromagnetic
nanoparticle arrays and nanodomain materials near the Curie
temperature in the limit of weak coupling between the grains. We
calculate the conductivity in the Ohmic and non-Ohmic regimes and
estimate the magnetoresistance jump in the resistivity at the
transition temperature. The results are applicable for many
emerging materials, including artificially self-assembled
nanoparticle arrays and a certain class of manganites, where
localization effects within the clusters can be neglected.
\end{abstract}

\maketitle

Arrays of ferromagnetic nanoparticles are becoming one of the
mainstreams of current mesoscopic
physics~\cite{Sun00,Black00,Zeng06,Majetich}. Not only ferromagnetic
granules promise to serve as logical units and memory storage
elements meeting elevated needs of emerging technologies, but also
offer an exemplary model system for investigation of disordered
magnets.  At the same time the model of weakly coupled nanoscale
ferromagnetic grains proved to be useful for understanding the
transport properties of doped manganite
systems~\cite{Dagotto01,Mathur03} that have intrinsic
inhomogeneities. Recent studies showed that above the Curie
temperature these materials possess a nanoscale ferromagnetic
cluster structure which to a large extend controls transport
in these systems~\cite{Fath,Moreo99,Mayr01,Moreo00}. This
defines an urgent quest for understanding and quantitative
description of electronic transport in ferromagnetic nanodomain
materials based on the model of nanogranular ferromagnets.

\begin{figure}[tbp] \hspace{-0.5cm}
\includegraphics[width=0.9\linewidth]{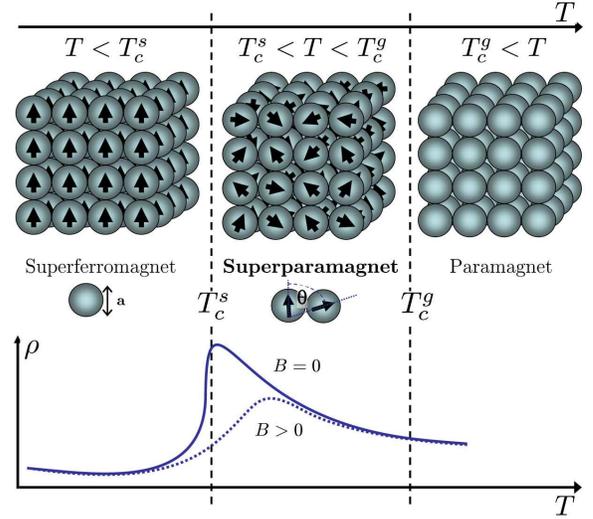}
\caption{Top: Sketch of a $3d$ granular system under consideration
showing the different states at different temperatures: For
$T<T_c^s$, where $T_c^s$ is the macroscopic Curie temperature of
the system, the ferromagnetic grains ({\it superspins}) form a
superferromagnet (SFM); for $ T_c^s < T < T_c^g$, where $T_c^g$ is
the Curie temperature for a single grain, the system is in a
superparamagnetic (SPM) state; and above $T_c^g$ the system shows
no magnetic order. In the SPM state the angle between two grains
(superspins) is denoted by $\theta$. Note, this is an idealized
picture, see~\cite{morph}. Bottom:
Schematic behavior of the resistivity, $\rho$, versus temperature,
$T$, in the different states in the absence ($B=0$) and presence
($B>0$) of a magnetic field $B$ aligned with the magnetization of
the SFM,  cf.~\cite{Zeng06,Chuang01}.\vspace{-0.7cm}} \label{array}
\end{figure}

In this \paper we investigate electronic transport properties of
arrays of ferromagnetic grains~\cite{morph} near the
ferromagnetic-paramagnetic transition, see Fig.~\ref{array}.
At low temperatures, $T < T_c^s$, the sample is in a  so called {\it
superferromagnetic} (SFM) state, see Fig.~\ref{array}, set up by
dipole-dipole interactions.  Near the \textit{macroscopic} Curie
temperature, $T_c^s$, thermal fluctuations destroy the macroscopic
ferromagnetic order.   At intermediate temperatures $T_c^s < T <
T_c^g$, where $T_c^g$ is the Curie temperature of a single grain,
the system is in a {\it superparamagnetic} (SPM) state where each
grain has its own magnetic moment while the global ferromagnetic
order is absent.  At even higher temperatures, $T > T_c^g$, the
ferromagnetic state within each grain is destroyed and the complete
sample is in a paramagnetic state.  We consider the model of
weakly interacting grains in which the sample Curie temperature is
much smaller than the Curie temperature of a single grain, $T_c^s
\ll T_c^g$.

We first focus on the SPM state, and discuss a $d-$dimensional
array ($d = 3,2$) of ferromagnetic grains taking into account
Coulomb interactions between electrons.  Granularity introduces
additional energy parameters apart from the two Curie
temperatures, $T_c^g$ and $T_c^s$: each nanoscale cluster is
characterized by (i) the charging energy $E_c=e^2/(\kappa a)$,
where $e$ is the electron charge, $\kappa$ the sample dielectric
constant, and $a$ the granule size, and (ii) the mean energy level
spacing $\delta$. The charging energy associated with nanoscale
ferromagnetic grains can be as large as several hundred
Kelvins~\cite{morph} and we require that $E_c/\delta \gg 1$. The
typical sample Curie temperature, $T_c^s$, of the arrays we
consider (and also of doped manganites) is in the range $(100 -
200)K$,~\cite{Zeng06,Chuang01,Uehara}; thus the temperature
interval $T_c^s < T < E_c$ is experimentally accessible. To
satisfy the last inequality, the size of a single ferromagnetic
grain, $a$, should be less than the critical size $a_c =
e^2/(T_c^s \kappa)$. The condition $E_c/\delta \gg 1$ defines the
lower limit for the grain size: $a_l = (\kappa /e^2
\nu)^{1/(D-1)}$, where $\nu=\nu_{\uparrow}+\nu_{\downarrow}$ is
the total density of states at the Fermi surface (DOS) with
$\nu_{\uparrow(\downarrow)}$ being the DOS for electrons with spin
up (down) and $D$ the grain dimensionality~\cite{delta}.

The internal conductance of a metallic grain is taken much larger
than the inter-grain tunneling  conductance, which is a standard
condition of granularity. The tunneling conductance is the main
parameter that controls macroscopic transport properties of the
sample~\cite{Beloborodov07}.  In consideration of applications to
experiments~\cite{Sun00,Zeng06,Tokura,Uehara} we restrict
ourselves to the case where the tunneling conductance is smaller
than the quantum conductance~\cite{weakcouple}. In the SPM state
the charge degrees are coupled with the spin degrees of freedom;
to reflect this connection, the tunneling conductance can be
written in a form $\tilde g_t(\theta) = g_t^{0}(1 + \Pi^2
\cos\theta)$,~\cite{Inoue96}, where $g_t^0$ is the tunneling
conductance in the paramagnetic state~\cite{conductance}; $\Pi =
(\nu_\uparrow - \nu_\downarrow) /(\nu_\uparrow + \nu_\downarrow)$
is the polarization factor of a ferromagnetic grain where
$\theta\in [0,\pi]$ is the angle between two superspins, see
Fig.~\ref{array}. The tunneling conductance, $\tilde g_t(\theta)$,
achieves its maximum value for parallel spins, $\theta = 0$
(corresponding to the SFM state). In general, the distribution of
angles $\theta$ is determined by some function $f(\theta)$ which
depends on temperature (and on external magnetic field): for
$T<T_c^s$, in the SFM state $f(\theta)$ is the
$\delta$-distribution and for high temperatures, $T \gg T_c^s$, it
is constant [an explicit expression for $f(\theta)$ is discussed
below Eq.~(\ref{sigmam})]. We denote averages over angles by
$\av{\ldots}_\theta\equiv\int_0^\pi d\theta\ldots f(\theta)$ with
$\av{1}_\theta=1$. Using this distribution we introduce the
averaged tunneling conductance:
\begin{equation}\label{conductance}
g_t(m^2)\equiv\av{\tilde g_t(\theta)}_\theta=g_t^0(1+\Pi^2m^2)\,,
\end{equation}
with the normalized ''magnetization'' $m^2=\av{\cos\theta}_\theta$,
e.g. $m^2=1$ in the SFM state and $m^2\to 0$ for high temperatures,
$T\gg T_c^s$. Note, that in general $m^2$ is not the normalized
(absolute value of the) magnetization of the sample since it only
takes into account the angle between two neighboring superspins in
the plane spanned by them (Fig.~\ref{array}). However, close to
$T_c^s$, we can expect $m^2\approx |{\bf M}(T)/M_s|^2$, where $M_s$
is the saturation value of the magnetization of the sample.
Below we first discuss the Ohmic transport near
$T_c^s$ and then summarize the main results for the
resistivity behavior in the non-Ohmic regime.


\paragraph{Ohmic transport}
To calculate the conductivity  for weakly coupled grains in the
presence of quenched disorder, we start with determining the total
probability for an electron to tunnel through $N$ grains $\tilde
P(\theta_1,\ldots,\theta_N) = \prod_{i=1}^N \tilde P_i(\theta_i)$,
where $\tilde P_i(\theta_i)$ denotes the probability for an
electron to tunnel through a single grain $i$ with an angle
difference $\theta_i$ of the magnetic moment to the previous
grain~\cite{Beloborodov07,validity}. The probability $\tilde
P(\theta_1,\ldots,\theta_N)$ has to be averaged over all angles in
order to obtain the total tunneling probability $ \P_{{\rm
total}}\equiv \langle \tilde
P(\theta_1,\ldots,\theta_N)\rangle_{\theta_1,\ldots,
\theta_N}=\prod_{i=1}^N \P_i(m^2)\,. $ The latter equality follows
from the fact that $\P_{{\rm total}}$ factorizes into the
individually averaged probabilities $\P_i(m^2)=\langle\tilde
P_i(\theta_i)\rangle_{\theta_i}$,~\cite{validity}. The mechanism
for electron propagation through an array of grains at low
temperatures is elastic and/or inelastic co-tunneling. The
corresponding probabilities in the limit of weak coupling between
the grains~\cite{weakcouple} are given by $\P_i^{\el}(m^2) \simeq
[ g_t(m^2) \delta ] / E_c$ and $\P_i^{\inel}(m^2) \simeq [
g_t(m^2) T^2] / E_c^2$, respectively,~\cite{Averin}. Assuming that
all probabilities $\P_i(m^2)$ are approximately the
same~\cite{morph} for each grain, $\P_i(m^2) = \P(m^2)$ for all
$i$, and expressing them in terms of the localization length
$\xi(m^2)$, defined by $\P(m^2)=\exp[-a/\xi(m^2)]$, we
obtain~\cite{Beloborodov07}
\begin{equation}
\label{xi} \xi^{\el} \simeq a/\ln [E_c/g_t(m^2) \delta],
\, \xi^{\inel} \simeq a/\ln [E_c^2/T^2
g_t(m^2)].
\end{equation}
Since the characteristic temperature we consider is of the order
of the Curie temperature, $T\sim T_c^s$, the dominant mechanism
for electron propagation is the inelastic
co-tunneling~\cite{prob}. Following Mott-Efros-Shklovskii's
theory~\cite{Mottbook,Efrosbook}, the conductivity can be written
as $\sigma(T,m^2) \sim g_t(m^2) \exp[{ -r/\xi(m^2) - e^2/(\kappa r
T)}]$, where the tunneling conductance $g_t(m^2)$ is given by
Eq.~(\ref{conductance}) and $r$ is the hopping distance. The first
term in the exponent accounts for electron tunneling and the
second term describes thermal activation necessary to overcome the
Coulomb correlation energy. Optimizing $\sigma(T,m^2)$ with
respect to the hopping length, $r=N\cdot a$, we obtain:
\begin{equation}
\label{sigmam} \sigma (T,m^2) \sim g_t^0(1 + \Pi^2 m^2)
\exp(-\sqrt{\T_0(m^2)/T}),
\end{equation}
with $\T_0(m^2) = T_0 [1 - (\xi_0/a)\ln(1 + \Pi^2 m^2)]$ being the
characteristic temperature scale. Here $T_0 \equiv
\T_0(m^2=0)=e^2/(\kappa \xi_0)$, where $\xi_0$ is the inelastic
localization length given in Eq.~(\ref{xi}) with the tunneling
conductance corresponding to the paramagnetic state,
$g_t(m^2=0)=g_t^0$. The minimal value of the resistivity in the
SFM state is determined by the minimal value of the energy scale
$\T_0^{\rm min} = T_0(1 - \ln2)$.

\begin{figure}[tbp] \hspace{-0.5cm}
\includegraphics[width=0.85\linewidth]{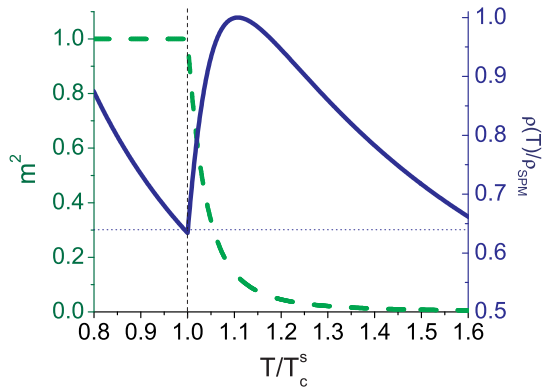}
\caption{Solid line, right axis: Plot of the normalized
resistivity $\rho(T,m^2)/\rho_{\SPM}$, inverse of the
conductivity $\sigma(T,m^2)$ in Eq.~(\ref{sigmam}), vs. temperature
for the following set of parameters: $\xi_0/a=1$, $T_0/T_c^s=10$,
and $\Pi^2=0.3$. Here $\rho_{\SPM}\equiv\rho(T_1,m_1^2)$ with $T_1>T_c^s$ being the temperature at which $\rho(T,m^2)$ is maximal and $m_1^2\equiv m^2(T_1)$. Dashed line, left axis: Plot of the
''magnetization'' $m^2$ versus temperature using an angular
distribution function defined in the text.\vspace{-0.7cm}} \label{m_rho}
\end{figure}

The behavior of the resistivity $\rho(T,m^2)$,  inverse of the
conductivity $\sigma(T,m^2)$ in Eq.~(\ref{sigmam}), and the magnetization $m^2(T)$ are
shown in Fig.~\ref{m_rho} for the following set of parameters:
$\xi_0/a=1$, $T_0/T_c^s=10$, and $\Pi^2=0.3$.
In order to describe typical experimental data~\cite{Chuang01}, the normalized distribution function $f(\theta)$ was chosen to ensure a sharp drop in the magnetization at $T_c^s$, $f[\alpha,x] = [\alpha/\arctan(1/\alpha)]
(x^2+\alpha^2)^{-1}$, with $x=\theta/\pi$ and
$\alpha(T)=10(T/T_c^s-1)$ for $T\geq T_c^s$ and
$\alpha(T<T_c^s)=0$ otherwise. The numerical constant in
$\alpha(T)$ was taken to produce a drop of $m^2$ in a temperature
region $\Delta T_c^s$ with $\Delta T_c^s/T_c^s\sim 0.1$.
Note, that the jump in the resistivity does not depend on the precise functional expression of $f(\theta)$; but it is sufficient that $m^2$ decays rapidly in the interval $\Delta T_c^s\ll T_c^s$.

For small polarization factors, $\Pi^2m^2 \ll 1$, the expression for
the energy scale $\T_0(m^2)$, can be written as $\T_0(m^2) = T_0 [1 -
(\xi_0/a)\Pi^2m^2]$. As a result, we obtain for the conductivity of an array of superspins  $\sigma(T,m^2) \sim [\, 1 +
\gamma_T\, \Pi^2 \, m^2 \, ]\exp(-\sqrt{T_0/T})$, where $\gamma_T$ is
a temperature dependent function.  For temperatures $ T_c^s \leq T
\ll T_0$ it is given  by $\gamma_T \simeq (\xi_0/2a) \,
\sqrt{T_0/T}$. The last expression can be written in terms of the optimal
hopping length, $r_{\opt} = \xi_0 \sqrt{T_0/T}$, as $\gamma_T \simeq
r_{\opt}/2a$. Since the hopping length $r_{\opt}$ depends on temperature
even within the SPM state, where the temperature $T$ satisfies the
inequality $ T_c^s < T < E_c$, one expects to observe two
different regimes: at temperatures $T_c^s < T< T^*$
the dominant mechanism for electron propagation is variable range
hopping [in this regime $r_{\opt} > a$ ], while for $ T^* < T < E_c$
electrons hop between the nearest neighbor grains only [$r_{\opt} \sim a
$]. The separating temperature $T^*$ can be estimated using the condition
$r_{\opt} \simeq a$ which gives $T^* \simeq T_0 \, (\xi_0/a)^2$, i.e.
$T^*\leq T_0$.

To calculate the magnitude of the jump value of the resistivity at the transition from the SPM to the SFM state in Fig.~\ref{m_rho}, we
introduce the dimensionless resistance ratio $\Delta \rho/\rho
\equiv [\rho_{\SPM} - \rho_{\SFM}]/ \rho_{\SPM}$, where
$\rho_{SFM}$ is calculated at temperature $T=T_c^s$, using the
inverse of Eq.~(\ref{sigmam}), for magnetization $m^2=1$;
$\rho_{\SPM}$ has to be evaluated at some temperature $T_1>T_c^s$ at
which $\rho(T,m^2)$ is maximal and magnetization $m_1^2\equiv
m^2(T_1)$. The temperature $T_1$ is of the order of $T_c^s+\Delta T_c^s$;
but since the magnetization drops quickly above $T_c^s$, i.e.
$\Delta T_c^s\ll T_c^s$, it is sufficient to take $\rho_{SPM}$ at
$T_1\sim T_c^s$ for calculating the jump. Using Eq.~(\ref{sigmam}) we obtain the following result:
\begin{equation}
\label{rho} \frac{\Delta \rho}{\rho} \simeq
1-\frac{1+\Pi^2m_1^2}{1+\Pi^2}\, {\rm
e}^{-\sqrt{\T_0(m_1^2)/T_c^s}+\sqrt{\T_0(1)/T_c^s}}\,.
\end{equation}
For small polarization factors, $\Pi^2 \ll 1$, this reduces to $\Delta \rho/\rho \simeq (\xi_0/2a) \,
\sqrt{T_0/T_c^s} \, \Pi^2 (1 - m_1^2)$, and can be expressed in terms of $r_{\opt}$ as
$\Delta\rho/\rho \sim \Pi^2 \, r_{\opt}(T_c^s)/a $. Therefore the larger the
hopping length, the bigger is the resistivity jump between the SPM and SFM
states.  From Eq.~(\ref{rho}) follows that the resistance ratio $\Delta \rho/\rho$ increases for small Curie
temperatures, $T_c^s \ll T_0$. We now estimate the jump magnitude in
Eq.~(\ref{rho}): Using the realistic values: $\Pi^2 =
0.3$, $m_1^2 = 0.1$, $\xi_0/a = 1$, $T_0/T_c^s = 10$, we obtain
$\Delta \rho/ \rho \simeq 0.4$, corresponding to a jump $\gamma =
60\%$, where $\gamma$ is defined by $\rho_{\SPM} = (1+\gamma) \, \rho_{\SFM}$. This estimate agrees with the plot in
Fig.~\ref{m_rho}.


\paragraph{Non-Ohmic regime}
So far we discussed the Ohmic regime in the absence of an
additional external electric field (or applied voltage) only.  In
the presence of an electric field $E$, the hopping conductivity in
the paramagnetic state is $\sigma \sim \exp \left[-r/\xi -
e^2/(\kappa r T) + eEr/T \right]$,~\cite{Shklovskii73}, with the
inelastic co-tunneling localization length $\xi^{\inel}=
a/\ln[E_c^2/g_t^0\, (T^2 +
(eEa)^2)]$,~\cite{Beloborodov07,Averin}. For sufficiently high
electric fields $E
> T/e\xi$ the tunneling term, $\exp(-r/\xi)$, in the conductivity is not important. As a result the optimal hopping
distance $r_{\opt}(E) \sim \xi \sqrt{E_{\xi}/E}$, with the characteristic electric field $E_{\xi} =
e/(\kappa \xi^2)$, and the resistivity $\rho \sim \exp[\, r_{\opt}(E)/\xi \, ]$ are temperature independent.

Including the magnetization dependent tunneling conductance
$g_t(m^2)$ in the above consideration, one finds that the
conductivity in the SPM state in the presence of a strong electric
field $\sigma(E,m^2)$ is given by Eq.~(\ref{sigmam}) with
$\T_0(m^2) \rightarrow \E_0(m^2)$ and $T\rightarrow E$, where
$\E_0(m^2) = E_0 [\, 1 - (2\xi_0/a) \ln(1 + \Pi^2m^2) \, ]$ is the
characteristic electric field, $E_0 = T_0/e\xi_0$, and $\xi_0 =
a/\ln[E_c^2/([eaE]^2g_t^0)]$. Equation~(\ref{sigmam}) with the
above substitutions holds for electric fields $T/\, [\, e \,
\xi(m^2)\,] < E < \E_0(m^2)$. The last inequality means that the
optimal hopping length $r_{\opt}(E)$ is larger than the size of a
single grain, $a$, while the first inequality ensures that the
electric field $E$ is still strong enough to cause non-Ohmic
behavior. Using typical values~\cite{Black00} for $a \approx
10$nm, $T_0 \approx 10^3$K, $\kappa \approx 3$, and temperature $T
\approx 10^2$K we estimate the window for electric fields as
$10^3$V/cm $< E < 10^5$V/cm. The resistance ratio $\Delta
\rho/\rho$ in the presence of a strong electric field is still
given by Eq.~(\ref{rho}) with the substitution $\T_0(m_1^2)
\rightarrow \E_0(m_1^2)$. For small polarizations, $\Pi^2 \ll 1$,
one obtains $\Delta \rho/\rho \simeq \, (\xi_0/a) \,
\sqrt{E_0/(T_c^s/e\xi_0)} \, \, \Pi^2 \, (1 - m_1^2)\,$.

\paragraph{Discussion} Past experimental studies of self-assembled
ferromagnetic arrays were dealing either with their thermodynamic
properties~\cite{Sun00,Black00,Zeng06,Majetich}, or with domain
wall motion~\cite{Petracic04}.  Investigations of the electronic
transport and magnetoresistance (MR) were mostly restricted to the
SPM state~\cite{Zhu99,Kakazei01}, where variable range hopping was
observed.  The crossover region near $T_c^s$ was only studied by
numerical methods in the context of manganite
systems~\cite{Dagotto01,Mayr01,Moreo00}. The resistivity
dependence below the Curie temperature $T_c^s$ presented in
Fig.~\ref{m_rho} is different from the schematic behavior shown in
Fig.~\ref{array} reflecting the experimental data on manganite
systems of Ref.~\cite{Chuang01}. In these materials the resistance
in the SFM state (below $T_c^s$) is close to the quantum
resistance~\cite{weakcouple} and therefore weakly depends on
temperature. In our consideration we assumed, based on the
experiments~\cite{Sun00,Zeng06,Uehara,Tokura}, that in the SFM
state the sample resistance is much larger than the quantum
resistance~\cite{weakcouple}, meaning that it exhibits variable
range hopping behavior and therefore is more sensitive to
temperature than the resistance of manganite systems.

Recently the nanoscale granularity in manganese oxides was directly
observed experimentally in
La$_{2-2x}$Sr$_{1+2x}$Mn$_2$O$_7$,~\cite{Ling00}. The cluster
structure in these perovskite materials is introduced by dopants,
creating the individual weakly coupled nanodomains.  To describe the
MR in these materials one has to take into account
electron localization within each cluster. This means that besides
the tunneling conductance, $g_t$, a finite grain conductance $g_0$
has to be considered as well. In this case the total conductance can
be written in the form $g(T,m^2) = g_0(T) \, g_t(m^2) / [g_0(T) +
g_t(m^2)]$. Below the Curie temperature $T_c^s$, the tunneling
conductance is small, $g_t(1) \ll g_0(T<T_c^s)$, such that
$g(T<T_c^s,1)=g_t(1)$, whereas above $T_c^s$, the grain conductance
$g_0(T)$ becomes small due to localization effects (e.g. Jahn-Teller
effect~\cite{Millis95}), leading to the formation of an insulation state, see e.g.~\cite{Chuang01}. In our  \paper we were assuming that $g_0
\gg g_t$ therefore the localization effects within each grain are
small. This situation is realized in e.g. La$_{1-x}${\it
A}$_x$MnO$_3$ ({\it A}= Sr, Ca),~\cite{Gu}.

The above considerations were carried out at zero external {\it magnetic
field} $B$. A finite field in ferromagnetic domain materials affects the
SFM-SPM transition and leads to a reduction of the peak in the
MR accompanied by a shift to higher temperatures
with increasing $B$, which is parallel to the magnetization of
the SFM at low $T$ [Fig.~\ref{array} (bottom);
Refs.~\cite{Chuang01,Zeng06}]. This means that
the distribution function $f(\theta)$ favors small angles near
$T_c^s$, i.e. the drop in $m^2$ is smeared out.

To summarize, we have investigated transport properties of
ferromagnetic nanoparticle arrays and nanodomain materials in the
limit of weak coupling between grains in the SPM and SFM states in
both Ohmic and non-Ohmic regimes. We have described the electron
transport near the Curie temperature $T_c^s$ in the artifically
self-assembled superspin arrays and discussed possible applications
of our results to a certain class of doped manganites, where
localization effects within the clusters can be neglected.  We
derived the magnitude of the jump in the resistivity at the
transition between the SPM and SFM states. We also discussed
the influence of the magnetic field on the jump amplitude and
the relation of our results to available experimental data.

\paragraph{Acknowledgements}
We thank Ken Gray, John Mitchell, Philippe Guyot-Sionnest,
Heinrich Jaeger, Wai Kwok, and Xiao-Min Lin for useful
discussions. This work was supported by the U.S. Department of
Energy Office of Science through contract No. DE-AC02-06CH11357.
A.~G. acknowledges support by the DFG through a research grant.
I.~B. was supported by the UC-ANL Consortium for Nanoscience
research.



\begin{thebibliography}{99}

\bibitem{Sun00} \vspace{-0.5cm} S.~Sun \etal, Science {\bf 287}, 1989 (2000).

\bibitem{Black00} C.~T.~Black \etal, Science {\bf 290}, 1131 (2000).

\bibitem{Zeng06} H.~Zeng \etal, Phys. Rev. B {\bf 73}, 020402(R) (2006).

\bibitem{Majetich} Y.~Ding and S.~A.~Majetich, Appl. Phys. Lett. {\bf 87},
022508 (2005); P.~Poddar \etal, Phys. Rev. B {\bf 68}, 214409 (2003).

\bibitem{Dagotto01} E.~Dagotto \etal, Phys. Rep.
{\bf 344}, 1 (2001).

\bibitem{Mathur03} N.~Mathur and P.~Littlewood, Phys. Today, {\bf 56} (1), 26 (2003).

\bibitem{Moreo99} S.~Yunoki \etal, Phys. Rev. Lett {\bf 80}, 845 (1998);
A.~Moreo \etal, Science {\bf 283}, 2034
(1999).

\bibitem{Fath} M.~F\"ath \etal, Science {\bf 285}, 1540 (1999).

\bibitem{Mayr01} M.~Mayr \etal, Phys. Rev. Lett. {\bf 86},
135 (2001).

\bibitem{Moreo00} A.~Moreo \etal, Phys. Rev. Lett. {\bf 84}, 5568 (2000).

\bibitem{morph} In this \paper (i) the grains are mesoscopically different
in the sense of disorder and their surface roughness on the atomic
scale and (ii) the tunneling conductances are of the same order.

\bibitem{Chuang01} Y.~D.~Chuang \etal, Science {\bf 292}, 1509
(2001); Y.~Moritomo \etal, Nature {\bf 380}, 141 (1996).

\bibitem{Uehara} M.~Uehara \etal, Nature {\bf 399}, 560 (1999).

\bibitem{delta} For a Curie temperature $T_c^s \sim 100 K$ and
dielectric constant $\kappa \sim 3$ the critical grain size is
$a_c \sim 50 {\rm nm}$. The mean energy level spacing $\delta$ for
a single cluster is expressed in terms of its volume $V$ and the
DOS $\nu$ at the Fermi surface as  $\delta =
(\nu V)^{-1}$. In this \paper we assume $T_c^s\gg\delta$.

\bibitem{Beloborodov07} I.~S.~Beloborodov \etal,
Rev. Mod. Phys. {\bf 79}, 469 (2007).

\bibitem{Tokura} Y.~Tokura \etal, J. Appl. Phys. {\bf 79}, 5288 (1996).

\bibitem{weakcouple} In references~\cite{Black00,Zeng06} the sample
resistance ($> 10$M$\Omega$) well exceeds the quantum resistance
[$h/(2e^2)=12.9$k$\Omega$].


\bibitem{Inoue96} J.~Inoue and S.~Maekawa, Phys. Rev. B {\bf 53},
R11927 (1996).

\bibitem{conductance} Tunneling conductance $g_t^0$ between
two clusters $i$ and $j$ is expressed in terms of the tunneling
matrix $t$ and DOS $\nu$: $g_t^0
= \langle |t|^2 \rangle \nu_i \nu_j$. $\langle \ldots
\rangle$ stands for disorder average.

\bibitem{validity} The total probability $\tilde P(\theta_1,\ldots,\theta_N)$
can be factorized as long as the Coulomb interaction is short
ranged, see~\cite{Beloborodov07}.


\bibitem{Averin} D.~A.~Averin and Yu.~V.~Nazarov, Phys. Rev.
Lett. \textbf{\ 65}, 2446 (1990).

\bibitem{prob} The two probabilities, $\P_i^{\el}$ and
$\P_i^{\inel}$, are of the same order for temperatures $T \simeq
\sqrt{E_c \delta}$. For temperatures larger than that, $\xi^{\inel}
> \xi^{\el}$, therefore inelastic co-tunneling is the dominant
transport mechanism.

\bibitem{Mottbook} N.~F.~Mott, {\it Metal-Insulator Transitions}, Taylor and Francis,
1990; N.~F.~Mott, Adv. Phys. \textbf{16}, 49 (1967).

\bibitem{Efrosbook} B.~I.~Shklovskii and A.~L.~Efros, {\it Electronic properties
of Doped Semiconductors, Springer, New York, 1988)};
L.~Efros and B.~I.~Shklovskii, J.~Phys.~C \textbf{8}, L49 (1975).

\bibitem{Shklovskii73} B.~I.~Shklovskii, Fiz. Tekh. Poluprovodn.
(S.-Petersburg) {\bf 6}, 2335 (1972) [Sov. Phys. Semicond. {\bf
6},1964 (1973)].

\bibitem{Petracic04} O.~Petracic, A.~Glatz, and W.~Kleemann, Phys. Rev. B {\bf 70}, 214432 (2004).

\bibitem{Zhu99} T.~Zhu and Y.~J.~Wang, Phys. Rev. B {\bf 60}, 11918 (1999).

\bibitem{Kakazei01} G.N.~Kakazei \etal, J. Appl. Phys. {\bf 90}, 4044 (2001); S.~Sankar \etal, Phys. Rev. B {\bf 62}, 14273 (2000).

\bibitem{Ling00} C.D.~Ling \etal, Phys. Rev. B {\bf 62}, 15096 (2000).
In this \paper the compound La$_{2-2x}$Sr$_{1+2x}$Mn$_2$O$_7$ was
studied, with doping level $x\in [0,1]$. For
$0.3\lesssim x\lesssim 0.4$ this maganese oxide system has a SFM to
SPM transition.


\bibitem{Millis95} A.~J.~Millis \etal,
Phys. Rev. Lett. \textbf{\ 74}, 5144 (1995).

\bibitem{Gu} K.~H.~Kim \etal, Phys. Rev. B \textbf{\ 55}, 4023 (1997).



\end{thebibliography}
\end {document}